\documentstyle[11pt,amssymb]{article}

\def\dalemb#1#2{{\vbox{\hrule height .#2pt
        \hbox{\vrule width.#2pt height#1pt \kern#1pt
                \vrule width.#2pt}
        \hrule height.#2pt}}}

\def\0{{\sst{(0)}}}
\def\1{{\sst{(1)}}}
\def\2{{\sst{(2)}}}
\def\3{{\sst{(3)}}}
\def\4{{\sst{(4)}}}
\def\5{{\sst{(5)}}}
\def\6{{\sst{(6)}}}
\def\7{{\sst{(7)}}}
\def\8{{\sst{(8)}}}
\def\n{{\sst{(n)}}}

\def\ep{\epsilon}

\let\a=\alpha \let\b=\beta \let\g=\gamma \let\d=\delta 
   \let\i=\iota 
 \let\m=\mu \let\n=\nu   \let\r=\rho
 \let\t=\tau    
\let\w=\omega  \let\D=\Delta

\def\nn{\nonumber} \def\bd{\begin{document}} \def\ed{\end{document}}
\def\ds{\documentstyle} \let\fr=\frac \let\bl=\bigl \let\br=\bigr
\let\Br=\Bigr \let\Bl=\Bigl 
\let\bm=\bibitem
\let\na=\nabla
\let\pa=\partial \let\ov=\overline 
\newcommand{\be}{\begin{equation}} 
\newcommand{\ee}{\end{equation}} 
\def\ba{\begin{array}}
\def\ea{\end{array}}
\def\ft#1#2{{\textstyle{{\scriptstyle #1}\over {\scriptstyle #2}}}}
\def\fft#1#2{{#1 \over #2}}
\def\del{\partial}
\def\sst#1{{\scriptscriptstyle #1}}
\def\oneone{\rlap 1\mkern4mu{\rm l}}
\def\ie{{\it i.e.\ }}
\def\via{{\it via}}
\def\semi{{\ltimes}}
\def\str{{\rm str}}
\def\jm{{\rm j}}
\def\im{{\rm i}}
\def\mapright#1{\smash{\mathop{-\!\!\!-\!\!\!-\!\!\!-\!\!\!-\!\!\!
             \longrightarrow}\limits^{#1}}}
\def\maprightt#1#2{\smash{\mathop{-\!\!\!-\!\!\!-\!\!\!-\!\!\!-\!\!\!
             \longrightarrow}\limits^{#1}_{#2}}}

\newcommand{\ho}[1]{$\, ^{#1}$}
\newcommand{\hoch}[1]{$\, ^{#1}$}
\newcommand{\bea}{\begin{eqnarray}} 
\newcommand{\eea}{\end{eqnarray}} 
\newcommand{\ra}{\rightarrow}
\newcommand{\lra}{\longrightarrow}
\newcommand{\Lra}{\Leftrightarrow}
\newcommand{\ap}{\alpha^\prime}
\newcommand{\bp}{\tilde \beta^\prime}
\newcommand{\tr}{{\rm tr} }
\newcommand{\Tr}{{\rm Tr} } 
\newcommand{\NP}{Nucl. Phys. }
\newcommand{\tamphys}{\it Center for Theoretical Physics\\
Texas A\&M University, College Station, Texas 77843}
\newcommand{\upenn}{\it Department of Physics and Astronomy\\
University of Pennsylvania, Philadelphia, Pennsylvania 19104}

\newcommand{\auth}{H. L\"u\hoch{\dagger1}, C.N. Pope\hoch{\ddagger2}
and T.A. Tran\hoch{\ddagger} }

\begin{document}
\begin{flushright}
\hfill{CTP TAMU-41/99}\\
\hfill{UPR-862-T}\\
\hfill{hep-th/9909203}\\
\hfill{September 1999}\\
\end{flushright}

\vspace{15pt}

\begin{center}
{ \large {\bf Five-dimensional $N=4$, $SU(2)\times U(1)$ Gauged Supergravity
from Type IIB}}

\vspace{15pt}
\auth

\vspace{15pt}

{\hoch{\dagger}\upenn}

\vspace{15pt}
{\hoch{\ddagger}\tamphys}

\vspace{40pt}

\underline{ABSTRACT}
\end{center}

We construct the complete and explicit non-linear Kaluza-Klein ansatz
for deriving the bosonic sector of $N=4$ $SU(2)\times U(1)$ gauged
five-dimensional supergravity from the reduction of type IIB
supergravity on $S^5$.  This provides the first complete example of
such an $S^5$ reduction that includes non-abelian gauge fields, and it
allows any bosonic solution of the five-dimensional $N=4$ gauged
theory to be embedded in $D=10$.

{\vfill\leftline{}\vfill
\vskip 5pt
\footnoterule
{\footnotesize \hoch{1} Research supported in part by DOE grant 
DE-FG02-95ER40893 \vskip -12pt} \vskip 14pt
{\footnotesize  \hoch{2} Research supported in part by DOE 
grant DE-FG03-95ER40917.\vskip  -12pt}}

\pagebreak
\setcounter{page}{1}

\section{Introduction}

    Recently, a new duality was conjectered, which relates
supergravities in anti-de Sitter backgrounds to superconformal field
theories on their boundaries \cite{malda,gkp,wit}.  Since the anti-de
Sitter backgrounds arise as solutions of lower-dimensional gauged
supergravities, the conjectured AdS/CFT correspondence has led to a
revival of interest in deriving these supergravities by Kaluza-Klein
reduction from the fundamental string theories and M-theory 
in $D=10$ and $D=11$.

    It has long been believed that many gauged supergravities arise from
the Kaluza-Klein reduction of eleven-dimensional supergravity or the type
IIA and IIB supergravities on certain spherical internal spaces.
At the linearised level, it is well established that the spectra of
the massless supermultiplets in the maximal gauged supergravities in
$D=4$ and $D=7$ coincide with those coming from appropriate truncations of the
$S^4$ and $S^7$ reductions of $D=11$ supergravity \cite{dp,PTV}.  Likewise,
the spectrum of maximal gauged supergravity in $D=5$ is known to arise
from a truncation of the $S^5$ reduction from type IIB supergravity
\cite{gunaydin,kim}.  

     Although these linearised results are rather easily established,
it is much harder to determine whether the truncations to the massless
supermultiplets are consistent as the full non-linear level.  The key
point of concern here is that once the non-linear interactions are
included, the possibility exists that in a full untruncated reduction,
source terms built purely from the fields of the massless multiplet
might arise in the field equations for the lower-dimensional massive
multiplets, making it inconsistent to set the massive fields to zero.
Indeed, it is not hard to establish that in a sphere reduction of a
generic higher-dimensional theory, there will definitely be such
couplings, making a consistent truncation to the massless sector
impossible.  Thus if the sphere reductions (with truncation to the
massless supermultiplet) in supergravities are to be consistent, it
must be because of remarkable special properties of these particular
theories.  General arguments suggesting that reductions to the
supergravity multiplet should always be consistent were constructed in
\cite{zilch}. 

    Early indications of such properties were found for the $S^7$
reduction of $D=11$ supergravity in \cite{dnpw}.  Subsequently, a
complete demonstration of the consistency in this case was presented
in \cite{deWitnicolai}, although the construction was rather implicit.
Recently, a somewhat less implicit construction was presented for the
easier case of the reduction ansatz for the $S^4$ reduction of $D=11$
supergravity, to give the maximal gauged supergravity in $D=7$
\cite{nvv}.  No analogous results have been obtained for the $S^5$
reduction of type IIB supergravity, however, and so for now the
consistency of this reduction remains conjectural.

   Although one would certainly like to know the complete reduction
ans\"atze in all cases, for many purposes it is in practice more
useful to have a fully explicit ansatz for a restricted subset of the
fields of the entire maximal massless multiplet.  For example, the
known charged anti-de Sitter black hole solutions of gauged
supergravities make use only of gauge fields in the maximal abelian
subgroup of the full gauge group.  Thus, for example, if one is
interested in oxidising these solutions back to the higher dimension
of the original parent supergravity, it would be sufficient to have a
consistent Kaluza-Klein reduction ansatz for the abelian truncation of
the full gauged theory.  Kaluza-Klein ans\"atze of this sort were
constructed in \cite{ten}, for the $S^7$ and $S^4$ reductions of
$D=11$ supergravity, and the $S^5$ reduction of type IIB supergravity.
In all cases fully non-linear bosonic ans\"atze were obtained.  In the
case of the $S^5$ reduction, the ansatz was fully consistent, while in
the reductions from $D=11$, the ans\"atze were fully consistent within
limited subspaces of solutions, including the charged AdS black holes.

    Another way of making the reductions more managable, while
maintaining full consistency, is to truncate from the maximal
supermultiplet to a smaller supergravity multiplet.  By doing this,
the full non-linear ansatz for the $S^4$ reduction from $D=11$ to
$N=1$ $SU(2)$-gauged supergravity in $D=7$ was obtained in \cite{lp1}.
Although in principle subsumed by the $N=2$ results in \cite{nvv}, in
practice the full explicitness of the $N=1$ results makes them more
transparent and usable in those cases where the $N=1$ truncation is
sufficient.

    In a similar vein, the bosonic sector of the full $N=2$
$SU(2)$-gauged supergravity in $D=6$ was recently obtained, as an
explicit consistent (local) $S^4$ reduction of the massive type IIA
theory \cite{clp1}.  A distinction in this case is that there {\it is}
no gauged supergravity in $D=6$ with the maximal ($N=4$) supersymmetry
that can occur in ungauged theories, and so the construction in
\cite{clp1} yields the largest gauged pure supergravity that exists in $D=6$.
 
   As far as the five-dimensional gauged supergravities are concerned,
no non-linear reduction ans\"atze other than the $U(1)^3$ maximal
abelian case in \cite{ten} have been obtained until recently.
Studying the states on the Coulomb branch of ${\cal N}=4$ super
Yang-Mills theory from the viewpoint of five-dimensional gauged
supergravity~\cite{fgpw} provides further indications of consistent
truncations from type IIB supergravity.  For example, in \cite{cglp}
non-linear reduction ans\"atze involving larger numbers of the scalar
fields of the maximal $D=4$, 5 and 7 theories were presented, although
without the inclusion of gauge fields.

   In this paper, we shall construct the complete and explicit
non-linear Kaluza-Klein reduction ans\"atze for the bosonic sector of
$N=4$, $SU(2)\times U(1)$ gauged supergravity in $D=5$, obtained from
the $S^5$ reduction of type IIB supergravity.  The $N=4$ gauged theory
in $D=5$ was constructed by L. Romans \cite{romans}; hereafter, we
shall refer to this model as the Romans theory.  Our ansatz allows us
to re-interpret any bosonic solution of the five-dimensional $N=4$
theory back in $D=10$.  The ansatz provides the first example of a
fully consistent non-linear Kaluza Klein reduction of type IIB
supergravity on $S^5$ that includes non-abelian gauge fields.

   A new feature of this $S^5$ reduction, which has not been
encountered in previous explicit constructions \cite{lp1,clp1}, is the
presence of higher-rank gauge potentials that transform non-trivially
under the gauge group.  Specifically, there are two 2-form potentials
in the $D=5$ theory, which form a charged doublet under the $U(1)$
factor of the gauge group.  

  The paper is organised as follows.   In section
2, we establish our notation and conventions, giving
Lagrangians and equations of motion for type IIB supergravity and the
$D=5,\ N=4\ SU(2)\times U(1)$ gauged supergravity.  In section 3, we 
obtain the reduction ans\"atze, and discuss some of the interesting
features.  After the conclusion, in section 4, we include an appendix
giving results for the Ricci tensor of the metric reduction ansatz.

\section{$D=10$ IIB supergravity and Romans' theory in $D=5$}

      The bosonic sector of ten dimensional type IIB supergravity
comprises the metric, a self-dual 5-form field strength, a scalar, an
axion, an R-R 3-form and an NS-NS 3-form field strength.  There is no
simple covariant Lagrangian for type IIB supergravity, on account of
the self-duality constraint for the 5-form.  However, one can write a
Lagrangian in which the 5-form is unconstrained, which must then be
accompanied by a self-duality condition which is imposed by hand at
the level of the equations of motion \cite{bergs}.  This type IIB 
Lagrangian is
\bea
{\cal L}^{IIB}_{10} &=& \hat R\, {\hat *\oneone} - \ft12
{\hat *d\hat{\phi}}\wedge d\hat{\phi} - \ft12
e^{2\hat{\phi}}\, 
{\hat *d\hat{\chi}}\wedge d\hat{\chi} - \ft14 {\hat *\hat H_\5}
\wedge \hat H_\5 \nn\\
& & -\ft12 e^{-\hat \phi} \, {\hat * \hat{F}^{2}_\3}\wedge\hat{F}^{2}_\3 -
\ft12 e^{\hat\phi}\, {\hat *\hat{F}^{1}_\3}\wedge \hat{F}^{1}_\3 - 
\ft12 \hat B_\4\wedge d\hat{A}_\2^{1}\wedge d\hat{A}_\2^{2}\ ,\label{d10lag}
\eea
where $\hat F^2_\3 = d\hat A^2_\2,\, \hat F^1_\3=d\hat
A^1_\2 - \hat{\chi}\, d\hat{A}^2_\2,\, \hat{H}_\5=d\hat{B}_\4 -
\ft12 \hat{A}_\2^1\wedge d\hat{A}_\2^2 + \ft12
\hat{A}_\2^2\wedge d\hat{A}_\2^1$, and we use hats to denote
ten-dimensional fields and the ten-dimensional Hodge dual $\hat
*$. The equations of motion following from the Lagrangian, together
with the self-duality condition, are
\bea
\hat R_{MN} &=& \ft12\pa_M\hat{\phi}\, \pa_N\hat{\phi} + \ft12 e^{2
\hat\phi}\, \pa_M\hat{\chi}\, \pa_N\hat{\chi} + \ft1{96} \hat H^2_{MN}\nn\\
& &+ \ft14 e^{\hat\phi}\, \Big((\hat{F}^{1}_\3)^2_{MN} -
\ft1{12}(\hat{F}^1_\3)^2 \hat{g}_{MN}\Big)\nn\\
& & + \ft14 e^{-\hat\phi}\,
\Big((\hat{F}^{2}_\3)^2_{MN} - \ft1{12}
(\hat{F}^2_\3)^2\hat{g}_{MN}\Big),
\label{10einstein}
\eea
\be
d{\hat * d\hat{\phi}} = - e^{2\hat{\phi}}\, {\hat * d\hat\chi}\wedge
d\hat{\chi} - \ft12 e^{\hat{\phi}}\, {\hat *\hat{F}_\3^1}\wedge
\hat{F}_\3^1 + \ft12 e^{-\hat{\phi}}\, {\hat * \hat{F}_\3^2}\wedge
\hat{F}_\3^2,
\label{10dilaton}
\ee
\be
d\Big(e^{2\hat{\phi}}\, {\hat * d\hat\chi}\Big) =
e^{\hat{\phi}}\, {\hat * \hat{F}_\3^1}\wedge \hat{F}_\3^2,
\label{10axion}
\ee
\be
d\Big(e^{\hat{\phi}}\, {\hat * \hat{F}_\3^1}\Big) = \hat{H}_\5\wedge
\hat{F}_\3^2\,,\qquad  
d\Big(e^{-\hat{\phi}}\, {\hat *\hat{F}_\3^2} -
\hat{\chi}\, e^{\hat{\phi}}\, {\hat *\hat{F}_\3^1}\Big) = -
\hat{H}_\5\wedge ( \hat{F}_\3^1 + \hat{\chi}\, \hat{F}_\3^2),
\label{103form}
\ee
\be
d({\hat *\hat{H}_\5}) = - \hat{F}_\3^1\wedge
\hat{F}_\3^2, \qquad  \hat{H}_\5 = {\hat *\hat{H}_\5}.
\label{105form}
\ee
The equations (\ref{10einstein})-(\ref{105form}) are precisely those 
which were found in \cite{schwarz}. 

   Turning now to the $N=4$ gauged supergravity in $D=5$
 \cite{romans}, it has a bosonic sector comprising the metric, a
 scalar,  the $SU(2)$ Yang-Mills  potentials $A^i_\m$, a $U(1)$ gauge
 potential $B_\m$, and two 2-form potentials
 $A^\a_{\m\n}$ which transform as a charged doublet under the $U(1)$.
The Lagrangian is given by
\bea
{\cal L} &=& R\, {*\oneone} - 3 X^{-2} {*dX}\wedge dX 
-\ft12 X^4\,  {*G_\2}\wedge G_\2
-\ft12 X^{-2}\, ({*F^i_\2}\wedge F^i_\2 + {*A^\a_\2}\wedge A^\a_\2)\nn\\
& & +\fr1{2g_1} \epsilon_{\a\b}\, A^\a_\2\wedge dA^\b_\2 - \ft12
A^\a_\2\wedge A_\2^\a\wedge B_\1 - \ft12 F^i_\2\wedge F^i_\2\wedge B_\1\nn\\
& &  + 2 g_2( g_2\, X^2 + 2\sqrt{2} g_1\, X^{-1})\, {*\oneone}\,,
\label{laromans}
\eea
where $X$ parameterises the scalar degree of freedom, and can be
written in terms of a canonically-normalised dilaton $\phi$ as 
$X=e^{-\ft1{\sqrt6}\, \phi}$.  The 2-form field strengths are given by
\be
F_\2^i = dA_\1^i + \ft12 g_2\, \ep^{ijk}\, A_\1^j\wedge
A_\1^k\,,\qquad 
G_\2= dB_\1\,.
\ee

    Without loss of generality we may set
$g_1=g$ and $g_2=\sqrt{2}g$, since the two independent gauge coupling
constants may be recovered by appropriate rescalings.  We also find it 
advantageous to adopt a complex notation for the two 2-form 
potentials, which form a charged doublet with respect to the
gauge field $B_\1$.  Thus we define
\be
A_\2 \equiv A_\2^1 + \im\, A_\2^2\,.\label{complexa}
\ee
We then find that the equations of motion are
\bea
d(X^{-1}\, {*dX}) &=& \ft13 X^4\,  {*G_\2}\wedge G_\2 - 
\ft16 X^{-2} \, ({*F^i_\2}\wedge F^i_\2 + {*{\bar A}_\2}\wedge A_\2)\nn\\
& & - \ft{4}{3}g^2\, (X^2 - X^{-1})\, {*\oneone} ,\nn\\
d(X^4\, {*G_\2}) &=& - \ft12 F^i_\2\wedge F^i_\2 - 
\ft12 {\bar A}_\2\wedge A_\2 ,\nn\\
d(X^{-2}\, {*F^i_\2}) &=& \sqrt{2}\, g \,
\epsilon^{ijk}\, X^{-2}\, {*F^j_\2}\wedge A^k_\1
- F^i_\2\wedge G_\2,\nn\\
X^{2}\, {*F_\3} &=& - \im \, g\, A_\2 \,,\nn\\
R_{\m\n} &=& 3 X^{-2}\,  \pa_\m X\, \pa_\n X - \ft{4}{3}g^2\,(X^2 +
2  X^{-1})\, g_{\m\n}\nn\\
& & + \ft12 X^4 \, (G_\m{}^\r G_{\n \r} -\ft16 g_{\m\n} \, 
G_\2^2) + \ft12 X^{-2}\, 
(F^{i\ \r}_\m F^{i}_{\n\r}  - \ft16 g_{\m\n}\, (F^i_\2)^2)\nn\\
& & + \ft12 X^{-2}\,  ({\bar A}_{(\m}{}^\r\,  A_{\n)\r} - \ft16
g_{\m\n}\, |A_\2|^2)\,,
\label{eomromans1}
\eea
where
\be
F_\3 =D A_\2 \equiv dA_\2  -\im\, g\, B_\1\wedge A_\2\,.\label{gauged}
\ee
The operator $D$ defined in this equation is the $U(1)$ gauge-covariant 
exterior derivative.
Note that the complex field $A_\2$ satisfies a first-order equation of
motion, of the kind referred to as ``odd-dimensional self-duality'' 
in \cite{tpv}.   

   As discussed in \cite{romans}, there are three inequivalent
theories, corresponding to different choices for the gauge coul,ings
in (\ref{laromans}): $N=4^0$ in which $g_2=0$; $N=4^+$ in which
$g_2=\sqrt{2}g_1$ and $N=4^-$ in which $g_2=-\sqrt{2}g_1$ (see also
\cite{wnc}).  The $N=4^+$ theory is obtained by truncating the gauged
$SO(6),\ N=8$ supergravity theory in five-dimensions, while the
$N=4^0$ and $N=4^-$ theories arise as truncations of non-compact $N=8$
supergravities. The equations of motion (\ref{eomromans1}) are
precisely those of the $N=4^+$ theory.  From the Lagrangian of the
Romans theory one might conclude that it is not possible to set the
$U(1)$ coupling constant $g_1$ to be zero. However, as was pointed out
in \cite{pcowdall}, after appropriate rescalings the limit can be
taken.

   In the next section, we shall
construct the ans\"atze that give this theory by reducing type IIB
supergravity on a 5-sphere.

\section{Reduction ans\"atze on the five-sphere}

   To construct the reduction ans\"atze, we 
follow a procedure similar to that used in \cite{lp1,clp1}.  We take as
our starting point the $U(1)^3$ abelian truncation obtained
in~\cite{ten}, which involved two independent scalar fields.  One can
perform a consistent truncation of this, in which two of the three $U(1)$
potentials are set equal, and at the same time one of the scalar
degrees of freedom in eliminated.  After doing so,
the metric on the internal 5-sphere takes the following form
\bea
d\omega_5^2 &=& X\, \Delta\, d\xi^2 + X^2\, s^2 ( d\tau -
g B_\1)^2\nn\\
& & + \ft14 X^{-1}\, c^2 \Big( 
d\theta^2 + \sin^2\theta\, d\phi^2 + (d\psi +\cos\theta\, d\phi 
- \sqrt{2} g\, A_\1)^2\Big)\ ,
\label{5met1}
\eea
where $X\equiv e^{-\fft1{\sqrt6}\, \phi}$, and we have defined
\be
c\equiv \cos\xi\,,\qquad s\equiv\sin\xi\,.
\ee
The function $\Delta$ is given by
\be
\Delta = X^{-2}\, s^2 + X\, c^2 .
\ee
At $X=1$, in the absence of the $U(1)$ gauge fields, the metric
(\ref{5met1}) describes a unit 5-sphere.  (The gauge field $A_\1$ 
appearing in (\ref{5met1}) is the one that comes from setting two
of the original $U(1)$ gauge fields in \cite{ten} equal.)

     We can now make a non-abelian generalisation of the
the 5-sphere metric ansatz, by introducing the three
left-invariant 1-forms $\sigma^i$ on the 3-sphere, which satisfy $d\sigma^i =
-\ft12 \ep_{ijk}\, \sigma^j\wedge \sigma^k$.  These can be written in
terms of the Euler angles as $\sigma_1+\im\, \sigma_2 = e^{-\im\psi}\,
(d\theta + \im\, \sin\theta\, d\phi)$, $\sigma_3=d\psi +
\cos\theta\, d\phi$.   Thus we are naturally led to generalise
(\ref{5met1}) to
\be
d\omega_5^2 = X\, \Delta\, d\xi^2 + X^2\, s^2 \Big(d\tau
- g B_\1 \Big)^2 + 
\ft14 X^{-1}\, c^2 \sum_{i=1}^3
(\sigma^i-\sqrt{2}g\, A_\1^i)^2\ .\label{5met2}
\ee
The abelian limit (\ref{5met1}) is recovered by setting the $i=1$ and
$i=2$ components of the $SU(2)$ gauge potentials to 
zero.

    By proceeding along these lines, we are eventually led 
to the following ans\"atze for the ten-dimensional metric,  
the 5-form field strength $\hat H_\5 \equiv \hat G_\5 + {\hat *\hat
G_\5}$, and the two 2-form potentials:
\bea
d\hat s^2_{10} &=& 
\Delta^{1/2}\, ds_5^2 +  g^{-2} \, X\,
\Delta^{1/2}\, d\xi^2 +
g^{-2}\Delta^{-1/2}\, X^2\, s^2\, \Big(d\tau - g\, B_\1\Big)^2 \nn\\
& &+ \ft14 g^{-2}\, \Delta^{-1/2}\, X^{-1}\,
c^2\, \sum_i (\sigma^i - \sqrt{2}g\, A_\1^i)^2\ ,\nn\\
\hat{G}_\5 &=& 2g\, U \,\varepsilon_5 - \frac{3 sc}{g} X^{-1}\, 
{*dX}\wedge d\xi + 
\frac{c^2}{8\sqrt{2}\, g^2} X^{-2}\, {*F^i_\2}\wedge h^j\wedge h^k
\, \varepsilon_{ijk} \nn\\
& &-\frac{sc}{2\sqrt{2}\,g^2} X^{-2}\, {*F^i_\2}\wedge h^i\wedge d\xi
- \frac{sc}{g^2} X^4\, {*G_\2}\wedge d\xi\wedge (d\tau - g B_\1),\nn\\
\hat{A}_\2 &\equiv& \hat A_\2^1 + \im\, \hat A_\2^2 =
 -\fr{s}{\sqrt{2}g}\, e^{-\im\,\tau}\, A_\2\,,\nn\\
\hat{\phi} &=& 0,\ \ \ \hat{\chi} = 0,
\label{metans}
\eea
where $h^i\equiv \sigma^i - \sqrt{2} g\, A_\1^i$, $U\equiv X^2\, c^2 +
X^{-1}\, s^2 + X^{-1}$, and $\ep_5$ is the volume form in the 
five-dimensional spacetime metric $ds_5^2$. 
Note that we have defined the complex 2-form potential $\hat A_\2\equiv
\hat A_\2^1 + \im\, \hat A_\2^2$ in the type IIB theory.
The ten-dimensional dilaton and the axion are constants,
which without loss of generality we have set to zero.  The $SU(2)$
Yang-Mills field strengths $F_\2^i$ are given by $F_\2^i = dA_\1^i +
\frac{1}{\sqrt{2}} g\, \ep_{ijk}\, A_\1^j\wedge A_\1^k$.  It should be
emphasised that all the fields $X$, $A_\2^\a$ and $A_\1^i$, and the
metric $ds_5^2$, appearing on the right-hand sides of (\ref{metans}),
are taken to depend only on the coordinates of the five-dimensional
spacetime.\footnote{Note that the metric reduction ansatz in
(\ref{metans}) has a fairly simply form, and fits in with the general
pattern of Kaluza-Klein metric ans\"atze, such as have been seen in
many previous examples.  (See \cite{ss,wna,nilsson} for some earlier
examples, and \cite{wnb,ten,lp1,clp1} for some more recent ones.)  It
is the determination of the reduction ans\"atze for the field
strengths that is the more difficult part of the problem, and since
the consistency of the reduction depends crucially on conspiracies
between the field strength and metric contributions, it is only when
the full ansatz is known that the consistency can be established.}
Note also that since the two 2-form potentials of the 5-dimensional
supergravity are charged under the $U(1)$ factor in the $SU(2)\times
U(1)$ gauge group, these fields appear in the
reduction ansatz with the appropriate $\tau$ dependence for charged
$U(1)$ harmonics.  

   The ten-dimensional Hodge dual of $\hat G_\5$ turns out to be
\bea
{\hat *\hat{G}_\5} &=& -\fr{sc^3}{4g^4}\, U\, 
\Delta^{-2}\, d\xi\wedge \omega\wedge
\epsilon_3 + \frac{3s^2\, c^4}{8g^4} \,X^{-2}\, \Delta^{-2}\, dX\wedge
\omega\wedge\epsilon_3 \nn\\
& &-\fr{s^2\, c^2}{8\sqrt{2}\, g^3}\, X^{-2}\, \Delta^{-1}\, 
F^i_\2\wedge h^j\wedge h^k\wedge
\omega\,  \epsilon_{ijk} + \fr{s\, c}{2\sqrt{2}\, g^3} \, 
F^i_\2\wedge h^i\wedge
d\xi\wedge \omega\nn\\
& & - \fr{c^4}{8g^3}\, X\, \Delta^{-1}\,  G_\2\wedge\epsilon_3 ,
\label{}
\eea
where $\w\equiv  d\t - g B_\1$, and $\ep_3 \equiv h^1\wedge h^2\wedge h^3$.  

   We must now verify that the ans\"atze (\ref{metans}) do indeed
yield a consistent reduction of the type IIB theory to the $N=4$
gauged supergravity in $D=5$.  Let us begin by noting that the Bianchi
identity for $\hat H_\5$, and the field equations for the NS-NS and
R-R 3-forms of the type IIB theory become, in the complex notation,
\be
d\hat H_\5 = \ft{\im}{2}\,
\hat{\bar F}_\3\wedge F_\3\,,\qquad 
d{\hat *\hat F_\3} = -\im\, \hat H_\5\wedge \hat F_\3\,.\label{complexeom}
\ee
We now find that the Bianchi identity for $\hat H_\5$
gives rise to the following five-dimensional equations:
\bea
d(X^{-1}\, {*dX}) &=& \ft43 g^2 \, (X^{-1}-X^2) {*\oneone} 
- \ft16 X^{-2}\, ({*F^i_\2}\wedge
F^i_\2 + {*{\bar A}_\2}\wedge A_\2)\nn\\
& &+ \fr13 X^4\, {*G_\2}\wedge G_\2,\nn\\
d(X^{-2}\, {*F^i_\2}) &=& \sqrt{2}\, g \,X^{-2}\,
\epsilon^{ijk}\, {*F^j_\2}\wedge A^k_\1 -
F^i_\2\wedge G_\2,\nn\\
d(X^4\, {*G_\2}) &=& -\ft12 F^i_\2\wedge F^i_\2 - 
\ft12 {\bar A}_\2\wedge A_\2\,.
\label{equation}
\eea
These are precisely equations of motion of the Romans theory.

     Plugging the ans\"atze~(\ref{metans}) into the equation of motion
for $\hat F_\3$ in (\ref{complexeom}) gives rise
to the five-dimensional first-order equation for $A_\2$ in 
(\ref{eomromans1}), and also the second-order equation
\be
D(X^{2}\, {* F_\3}) = g^2\, X^{-2}\, {*A_\2}\,,\label{secondorder}
\ee
where the $U(1)$ gauge-covariant exterior derivative $D$ is as defined
in (\ref{gauged}).  This second order equation of motion is in fact
implied by the first order equation.

       With our ans\"atze, the ten-dimensional equations of motion for
the dilaton and axion are automatically satisfied. The final step is
to substitute the ans\"atze into the ten-dimensional Einstein equation
(\ref{10einstein}).  There are ten independent types of components of
the Ricci tensor, of which eight are non-vanishing, and are given in
the appendix.  (The components of the Einstein equation
(\ref{10einstein}) associated with the two vanishing ones,
$\hat{R}_{56}$ and $\hat{R}_{5i}$, just give identities 
$0=0$.)  The $\hat{R}_{\a
5}$ and $\hat{R}_{5i}$ components simply yield identities of the type
$Z=Z$ in (\ref{10einstein}), while $\hat{R}_{\a 6}$ gives an equation
of motion for the $U(1)$ field strength $G_\2$ which coincides with
(\ref{eomromans1}). The $\hat{R}_{55}$, $\hat{R}_{66}$ and
$\hat{R}_{ij}$ components reproduce the scalar equation of motion in
(\ref{eomromans1}).  Finally, the $\hat{R}_{\a\b}$ components yield
the five-dimensional Einstein equations, while $\hat{R}_{\a\i}$ give
the equations of motion for the $SU(2)$ gauge field strengths.  Thus
the full consistency of the reduction ansatz (\ref{metans}) is
established.

\section{Conclusion}
 
     In this paper, we have constructed the explicit and complete
all-orders ansatz for obtaining the bosonic sector of $SU(2)\times
U(1)$ gauged $N=4$ supergravity in $D=5$, by Kaluza-Klein reduction on
$S^5$ from type IIB supergravity.  This is the largest supersymmetric
$S^5$ reduction that has been constructed.  This is a particularly
significant case from the point of view of the AdS/CFT correspondence,
since the $D=5$ gauged supergravities coming from type IIB have
AdS$_5$ solutions whose boundary CFT's are four-dimensional.  In this
embedding, we showed that the ten-dimensional moduli, parameterised by
the dilaton and the axion, are fixed.

    Our results for this consistent $N=4$ reduction lend further
credence to the conjecture that the $SO(6)$ gauged $N=8$ supergravity
in $D=5$ should also arise as a consistent $S^5$ reduction.  The $N=8$
reduction can be expected to be extremely complicated, and so our
simpler $N=4$ truncation could prove to be very useful in
circumstances where only the $N=4$ subset of fields are excited.  For
example, our reduction ansatz allows any bosonic solution of the $N=4$
gauged theory in five dimensions to be oxidised back to an exact
solution in type IIB supergravity.  From the standpoint of the AdS/CFT
conjecture, the fact that there is a consistent reduction ansatz means
that the contributions of the five-dimensional fields in massive
supergravity multiplets can be ignored when computing correlation
functions in the conformal field theory on the boundary.

\section{Appendix}
In this appendix we present the vielbeins, spin connection, and Ricci
tensor components for the metric (\ref{metans}). The vielbeins are
taken to be
\bea
\hat{e}^\a &=& \Delta^{1/4} e^\a,\ \ \ 
\hat{e}^5 = \fr1{g}\, \Delta^{1/4}\, X^{1/2}\, d\xi ,\nn\\
\hat{e}^6 &=& \fr{s}{g}\Delta^{-1/4}\,  X \, (d\t - g B_\1),\ \ \ 
\hat{e}^{\underline{i}} = \fr{c}{2g}\Delta^{-1/4}\, X^{-1/2}\, h^i.
\eea
 From these we find that the connection 1-form components are given by
\bea
\hat{\w}_{\a\b} &=& \w_{\a\b} - \ft14 (c^2-2 X^{-3} s^2)\D^{-5/4}
\, (\pa_\a X \eta_{\b\g} - \pa_\b X \, \eta_{\a\g})\, \hat{e}^\g\nn\\
 & & + \fr{s}{2}\, X\, \D^{-3/4}\, G_{\a\b}\, \hat{e}^6 +
\fr{c}{2\sqrt{2}}\, X^{-1/2}\, \D^{-3/4}\, F^i_{\a\b}\, 
\hat{e}^{\underline{i}},\nn\\
\hat{\w}_{\a 5} &=& \fr{g\, s\, c}{2}\, X^{-1/2}\, 
(X^{-2}-X)\D^{-5/4}\, \hat{e}^\a -
\fr{3 c^2}{4}\, \D^{-5/4}\, \pa_\a X \, \hat{e}^5,\nn\\
\hat{\w}_{\a 6} &=& -\ft34 \D^{-5/4}\, (c^2 + 2 X^{-3}\, s^2)\, \pa_\a
X \, \hat{e}^6 + \fr{s}{2}\, X\, \D^{-3/4}\, G_{\a\b} \, \hat{e}^\b,\nn\\
\hat{\w}_{\a i} &=& \fr{3c^2}{4}\, \D^{-5/4}\, \pa_\a X\, 
\hat{e}^{\underline{i}} + \fr{c}{2\sqrt{2}}\, X^{-1/2}\, \D^{-3/4}
\, F^i_{\a\b} \, \hat{e}^\b ,\nn\\
\hat{\w}_{56} &=& - \fr{gc}{2s}\, X^{-1/2}\, ( \D + X)\, \D^{-5/4}\, 
\hat{e}^6,\nn\\
\hat{\w}_{5 i} &=& \fr{gs}{2c}\, X^{-1/2}\, ( \D + X^{-2})
\D^{-5/4}\, \hat{e}^{\underline{i}} ,\nn\\
\hat{\w}_{ij} &=& -
\fr{g}{c}\, X^{1/2}\, \D^{1/4}\, \varepsilon^{ijk}\, \hat{e}^{\underline{k}} -
\sqrt{2} \, g \, \D^{-1/4} \, \varepsilon^{ijk}\,  A^k_\a \, \hat{e}^\a ,
\label{vielbein1}
\eea
and the other components are zero. The non-vanishing Ricci tensor 
components in the vielbein basis are
\bea
\hat{R}_{\a\b} &=& \D^{-1/2}\, 
R_{\a\b} - \ft14 (c^2 - 2 X^{-3}\,  s^2) \D^{-3/2}
\, \Box X \, \eta_{\a\b}\nn\\
& & + \ft14 \D^{-5/2}\,  (c^4 - 10 X^{-3} \, s^2 \, c^2 - 2
X^{-6}\,  s^4)\, (\pa X)^2 \, \eta_{\a\b}\nn\\
& & -\ft12 (6 c^4 + 3 X^{-3}\,  s^2 \, c^2 + 6 X^{-6}\,  s^4) \pa_\a
\, X\pa_\b X\nn\\
& & + g^2\,  \D^{-5/2} \, (X^{-3} - 1)(2 X^{-2} \, s^4 + 2 X \, c^2 \,
s^2 - X^2 \, c^2)\, \eta_{\a\b}\nn\\
& & - \ft12 s^2\,  X^2 \, \D^{-3/2}\,  G_{\a\r}G_\b{}^\r - \ft14
c^2\, X^{-1}\, \D^{-3/2} \, F^i_{\a\r}\, F^{i\ \r}_{\b}\nn\\
\hat{R}_{\a 5} &=& -3g\, s\, c\, X^{-1/2}\, \D^{-5/2}\, U\, \pa_\a X\,,\nn\\
\hat{R}_{\a 6} &=& -\fr{s}{2}\, X\, \D^{-1}\,\na^\b G_{\a\b} - \fr{s}{2}
\,\D^{-2}\, (\D + 3 X^{-2}\, s^2) (\pa^\b X)\,  G_{\a\b}\,,\nn\\
\hat R_{\a i} &=& -\ft1{2\sqrt2}\, c\, X^{-1/2}\, \D^{-1}\, D^\b\,
F^i_{\a\b} + \ft1{2\sqrt2}\, X^{-1/2}\, \D^{-2}\, (2c^2- X^{-3}\,
s^2)\, \del^\b X\, F^i_{\a\b},\nn\\
\hat{R}_{55} &=& -\ft34 c^2\, \D^{-3/2}\, \Box X + \ft34 c^2\, \D^{-5/2}
\,(c^2 - 2 X^{-3}\, s^2)\, (\pa X)^2 \nn\\
& &+ g^2 \,\D^{-5/2}\, ( X^{-2} + 3 \D),\nn\\
\hat{R}_{66} &=& - \ft34 (c^2 + 2 X^{-3}\, s^2) \,\D^{-3/2} \,\Box X + \ft14
s^2 \,X^2 \,\D^{-3/2} \,G_\2^2 \nn\\
& & + \ft34 \D^{-5/2} \,( c^4 + 2 X^{-6} \,s^4 + 6 X^{-3} \,s^2 \,c^2) 
(\pa X)^2\nn\\
& & + g^2 \,\D^{-5/2} \,( 2 X \,c^4 + X\, c^2 + 2 X^{-2} + 4 X^{-2}\, 
c^2 \,s^2 + 2 X^{-5}\, s^4 - X^{-2}\, c^4 ),\nn\\
\hat R_{6i} &=& \fft{s\, c}{4\sqrt2}\, X^{1/2}\, \D^{-3/2}\, F^i_{\a\b}\, 
G^{\a\b}\,,\nn\\
\hat{R}_{ij} &=& \fr{3c^2}{4}\, \D^{-3/2} \,\Box X \, \d_{ij} - 
\fr{3c^2}{4}\, \D^{-5/2}\, (c^2 - 2 X^{-3}\, s^2) (\pa X)^2 \,\d_{ij} + 
\ft18 c^2\, \D^{-3/2} \,F^i_{\a\b}\,F^{j\ \a\b}\nn\\
& & + 
g^2\, \D^{-5/2} \,( 2 X^2 \,\D^2 + X\, U) \,\d_{ij}\,,
\label{ricci}
\eea
where $D^\a\, F^i_{\a\b} \equiv \nabla^\a\, F^i_{\a\b} +
\ft1{\sqrt2}\, g\, \ep^{ijk}\, F^j_{\a\b}\, A^{k\, \a}$.

   The scalar curvature is given by 
\bea
\hat{R} &=& \D^{-1/2}\, R - \ft12 (c^2 - 2 X^{-3} \,s^2)\, \D^{-3/2}\, \Box X
- \ft12 (5 X^{-1} \,c^2 + 8 X^{-4} \,s^2) \,\D^{-3/2} \,(\pa X)^2\nn\\
& & + g^2\, \D^{-3/2}\, (14 - 6 c^2 + 6 X^3\, c^2 + 12 X^{-3} \, s^2)\nn\\
& & - 
\fr{s^2}{4}\, X^2 \,\D^{-3/2} \,G_\2^2 - \fr{c^2}{8}\, X^{-1}\, \D^{-3/2}
\, (F_\2^i)^2
\eea


\begin{thebibliography}{99}
\bibitem{malda} J. Maldacena, {\sl The large $N$ limit of
superconformal field theories and supergravity},
Adv. Theor. Math. Phys. {\bf 2} (1998) 231, hep-th/9711200.

\bibitem{gkp} S.S. Gubser, I.R. Klebanov and A.M. Polyakov, {\sl Gauge
theory correlators from non-critical string theory}, Phys. Lett. {\bf
B428} (1998) 105, hep-th/9802109.

\bibitem{wit} E. Witten, {\sl Anti-de Sitter space and holography},
Adv. Theor. Math. Phys. {\bf 2} (1998) 253, hep-th/9802150.

\bm{dp} M.J. Duff and C.N. Pope, {\sl Kaluza-Klein supergravity
and the seven sphere}, in: Supersymmetry and supergravity 82, eds.
S. Ferrara, J.G. Taylor and P. van Nieuwenhuizen (World Scientific,
Singapore, 1983).

\bibitem{PTV} K. Pilch, P. K. Townsend and P. van Nieuwenhuizen, {\sl
Compactification of $d=11$ supergravity on $S^{4}$ (or $11=7+4$,
too)}, Nucl. Phys. {\bf B242} (1984) 377.

\bm{gunaydin}M. G\"unaydin and N. Marcus, {\sl The spectrum of the
$S^5$ compactification of the $N=2, D=10$ supergravity and the unitary
supermultiplet}, Class. \& Quantum Grav. {\bf 2} (1985) L11.

\bm{kim} H.J. Kim, L.J. Romans, and P. van Nieuwenhuizen, {\sl Mass
spectrum of ten dimensional $N=2$ supergravity on $S^5$},
Phys. Rev. {\bf D32} (1985) 389.


\bm{zilch} C.N. Pope and K.S. Stelle, {\sl  Zilch currents, supersymmetry
and Kaluza-Klein consistency}, Phys. Lett. {\bf B198} (1987) 151.

\bm{dnpw} M.J. Duff, B.E.W. Nilsson, C.N. Pope and N.P. Warner, {\sl
On the consistency of the Kaluza-Klein ansatz}. Phys. Lett. {\bf B149}
(1984) 90.


\bibitem{deWitnicolai} B. de Wit and H. Nicolai, {\sl The consistency
of the $S^{7}$ truncation in $D=11$ supergravity}, Nucl. Phys. B281
(1987) 211.

\bm{ten} M. Cveti\v{c}, M.J. Duff, P. Hoxha, J.T. Liu, H. L\"u,
J.X. Lu, R. Martinez-Acosta, C.N. Pope, H. Sati and T.A. Tran, {\sl
Embedding AdS black holes in ten and eleven dimensions},
hep-th/9903214, to appear in Nucl. Phys. {\bf B}.

\bm{nvv} H. Nastase, D. Vaman and P. van Nieuwenhuizen, {\sl
Consistent nonlinear KK reduction of 11d supergravity on $AdS_7\times
S_4$ and self-duality in odd dimensions}, hep-th/9905075.

\bm{lp1} H. L\"u and C.N. Pope, {\sl Exact embedding of $N=1,\ D=7$ gauged
supergravity in $D=11$,} hep-th/9906168, to appear in Phys. Lett. {\bf B}.

\bm{clp1} M. Cveti\v{c}, H. L\"u and C.N. Pope, {\sl Gauged
six-dimensional supergravity from massive type IIA,} hep-th/9906221.

\bm{fgpw} D.Z. Freedman, S.S. Gubser, K. Pilch and N.P. Warner, {\sl
Continuous distributions of D3-branes and gauged supergravity},
hep-th/9906194. 

\bm{cglp} M. Cveti\v{c}, S.S. Gubser, H. L\"u and C.N. Pope, {\sl
Symmetric potentials of gauged supergravities in diverse dimensions
and Coulomb branch of gauge theories}, hep-th/9909121.

\bm{romans} L.J. Romans, {\sl Gauged $N=4$ supergravities in five
dimensions and their magnetovac backgrounds,} Nucl. Phys. {\bf B267}
(1986) 433. 

\bm{bergs} E. Bergshoeff, C.M. Hull and T. Ortin, {\sl Duality in the
type II superstring effective action}, Nucl. Phys. {\bf B451} (1995)
547, hep-th/9504081.

\bm{schwarz} J.H. Schwarz, {\sl Covariant field equations of the chiral
$N=2, D=10$ supergravity}, Nucl. Phys. {\bf B226} (1983) 269.

\bm{tpv} P.K. Townsend, K. Pilch and P. van Nieuwenhuizen, {\sl
Selfduality in odd dimensions}, Phys. Lett. {\bf B136} (1984) 38,
Addendum-ibid. {\bf B137} (1984) 443.

\bm{wnc} M. G\"unaydin, L.J. Romans and N.P. Warner, {\sl Compact and
non-compact gauged supergravity theories in five dimensions},
Nucl. Phys. {\bf B272} (1986) 598.

\bm{pcowdall} P.M. Cowdall, {\sl On gauged maximal supergravity in six 
dimensions,} hep-th/9810041.

\bm{ss} Abdus Salam and J. Strathdee, {On Kaluza-Klein theory}, Ann.
Phys. {\bf 141} (1982) 316.

\bm{wna} B. de Wit, H. Nicolai and N.P. Warner, {\sl The embedding of
gauged $N=8$ supergravity into $D=11$ supergravity}, Nucl. Phys. {\bf
B255} (1985) 29.
 
\bm{nilsson} B.E.W. Nilsson, {\sl On the embedding of $D=4$ $N=8$
gauged supergravity in $D=11$ $N=1$ supergravity}, Phys. Lett. {\bf
B155} (1985) 54.

\bm{wnb} A. Khavaev, K. Pilch and N.P. Warner, {New vacua of gauged
$N=8$ supergravity in five dimensions}, hep-th/9812035.


\end{thebibliography}
\end{document}